\documentclass[aps,prl,twocolumn,nofootinbib,amsmath,superscriptaddress,floatfix]{revtex4-1}
\usepackage{graphicx}
\usepackage[T1]{fontenc}
\usepackage{textcomp}
\usepackage{txfonts}
\usepackage{bm}

\newcommand{\femnpt}{Fe$_{1-x}$Mn$_x$Pt}
\makeatletter
\newcommand*{\balancecolsandclearpage}{%
  \close@column@grid
  \clearpage
  \twocolumngrid
}
\makeatother

\begin{document}

\title{\emph{Ab initio} construction of magnetic phase diagrams in alloys: The case of Fe$_{1-x}$Mn$_x$Pt}

\author{B. S. Pujari}
\affiliation{Department of Physics and Astronomy and Nebraska Center for Materials and Nanoscience, University of Nebraska-Lincoln, Lincoln, Nebraska 68588, USA}
\affiliation{Centre for Modeling and Simulation, Savitribai Phule University of Pune, Ganeshkhind, Pune 411007, India}

\author{P. Larson}
\altaffiliation[Present address: ]{Viterbo University, La Crosse, Wisconsin 54601, USA}

\affiliation{Department of Physics and Astronomy and Nebraska Center for Materials and Nanoscience, University of Nebraska-Lincoln, Lincoln, Nebraska 68588, USA}

\author{V. P. Antropov}
\affiliation{Ames Laboratory, Ames, Iowa 50011, USA}

\author{K. D. Belashchenko}
\affiliation{Department of Physics and Astronomy and Nebraska Center for Materials and Nanoscience, University of Nebraska-Lincoln, Lincoln, Nebraska 68588, USA}
\date{\today}

\begin{abstract}
A first-principles approach to the construction of concentration-temperature magnetic phase diagrams of metallic alloys is presented.  The method employs self-consistent total energy calculations based on the coherent potential approximation for partially ordered and noncollinear magnetic states and is able to account for competing interactions and multiple magnetic phases. Application to the  Fe$_{1-x}$Mn$_x$Pt ``magnetic chameleon'' system yields the sequence of magnetic phases at $T=0$ and the $c$-$T$ magnetic phase diagram in good agreement with experiment, and a new low-temperature phase is predicted at the Mn-rich end. The importance of non-Heisenberg interactions for the description of the magnetic phase diagram is demonstrated.
\end{abstract}

\maketitle

Magnetic substitutional alloys are often found to excel in applications \cite{Kronmuller,Liu}, because alloying broadens the parameter space for tuning the desired properties. However, wide tunability, combined with the need to target certain operating temperature ranges, presents a challenge for empirical materials design. Competing magnetic interactions in alloys can produce complicated magnetic phase diagrams (MPD) with multiple magnetic phases \cite{Fruchart,Shimizu,Menshikov-rev,Tokura}. Understanding of the $c$-$T$ MPD's is therefore essential for the development of advanced magnetic materials.

Some MPD's can be computed using the Heisenberg model with empirical or calculated exchange parameters combined with the mean-field approximation (MFA) \cite{Medvedev,Kaphle} Monte-Carlo simulations \cite{Rusz,Ruban,Pal,Buchel}, or spin-fluctuation theory \cite{Sandr}. However, many systems are not adequately described by the Heisenberg model. In metallic alloys the interaction parameters are sensitive to the electronic structure and population and thereby to the content of the alloy \cite{Rusz,Pal,Kaphle} and to the degree of spin disorder \cite{Boettcher}. To avoid the limitations of the Heisenberg model, one can use first-principles spin-dynamics simulations \cite{SD} or construct a generalized spin Hamiltonian to map the adiabatic energy surface \cite{Wysocki,Glasbrenner} for use in thermodynamic calculations. The energies of disordered spin configurations can also be obtained using the disordered local moment (DLM) model \cite{Oguchi,DLM}, where the spin-rotational averaging is done in the coherent potential approximation (CPA). While all these approaches fail in strongly itinerant magnets, they are applicable when the spin moments do not vary by more than 10-20\% in different spin configurations. We restrict ourselves to such systems here.

First-principles spin-dynamics and the construction of a microscopic generalized spin Hamiltonian are computationally very demanding and unfeasible for most systems of practical interest. We have developed \cite{APS} an alternative approach, in which self-consistent DLM and noncollinear CPA calculations are used to construct a Ginzburg-Landau-type total-energy functional expressed through a small number of relevant magnetic order parameters. Combined with the MFA expression for the magnetic entropy, this method provides the variational free energy. A similar scheme was used to describe the phase transitions in FeRh \cite{Staunton}. Here we show, using the \femnpt\ disordered alloy system as a test case, that this efficient approach is sufficiently powerful to explain and refine a complicated MPD, not only reproducing the five known magnetic phases but also predicting another, hidden low-temperature phase in this system.

\femnpt\ alloys are of interest for ultrahigh-density magnetic recording and medical applications \cite{Gruner}. Their structural ordering is of the L1$_0$ type in the fcc sublattice, with (001) layers of Pt alternating with disordered Fe/Mn layers. Neutron diffraction measurements revealed three collinear and two noncollinear phases \cite{Menshikov}. The collinear phases are the ferromagnetic (F) phase at the Fe-rich end, the C-type antiferromagnetic phase at the Mn-rich end, and the G-type antiferromagnetic phase in the middle of the diagram. The corresponding ordering wave vectors are $\mathbf{Q}_F=(0,0,0)$, $\mathbf{Q}_C=(1,0,0)$, and $\mathbf{Q}_G=(1,0,1/2)$ in units of $2\pi/a$ (or $2\pi/c$ for the $z$ component). The transitions between the collinear phases occur through intermediate $2Q$ phases combining the corresponding orderings for two orthogonal spin components (see Supplemental Material \cite{sm} for an illustration).

The samples studied in Ref.\ \onlinecite{Menshikov} show a high degree of L1$_0$ order at all concentrations, which is consistent with the fact that magnetic ordering occurs well below the structural ordering temperatures in these alloys. Therefore, as a practical simplification we assumed perfect L1$_0$ ordering and complete disorder of Fe and Mn atoms within their own sublattice. The method can also be applied to alloys with partial chemical ordering and in principle allows one to study the coupling between magnetic and chemical order parameters \cite{Bieber}.

To construct the energy functional, we have extended our CPA implementation \cite{Fe8N,Fe2B} based on the tight-binding linear muffin-tin orbital formalism \cite{Turek} by special features designed to describe complicated magnetic states. First, we implemented the vector DLM (VDLM) model, in which partially ordered magnetic states are specified by the Curie-Weiss distribution functions $p_{i\mu}(\theta)\propto\exp(\alpha_{i\mu}\cos\theta)$, where $i$ is the lattice site and $\mu$ the component index; $\alpha_{i\mu}$ are regarded as variational parameters. This formulation is suitable for systems with axial spin symmetry, i.\ e.\ those with collinear magnetic order. Different spin moment orientations are treated as different alloy components in the CPA formalism. The integral over the azimuthal angle in the CPA equations is taken analytically, while the $\theta$ dependence is discretized using the 16-point Gauss-Legendre quadrature. The potentials for all atoms are determined by embedding the CPA self-consistency loop into the density-functional theory (DFT) charge iteration. To enforce magnetic self-consistency, constraining transverse magnetic fields \cite{Dederichs,Stocks} are introduced for each orientation of the spin moment and determined self-consistently. Local density approximation is used for exchange and correlation.

The second feature extends CPA to the noncollinear case, in which the orientations of the spin moments of different components on the same lattice site can be different. This method is suitable, in particular, for the studies of $2Q$ structures appearing in the \femnpt\ system. Self-consistent constraining fields are also used in these calculations. Both VDLM and noncollinear CPA calculations yield the DFT total energy.

\begin{figure*}[htb]
\includegraphics[height=0.21\textwidth]{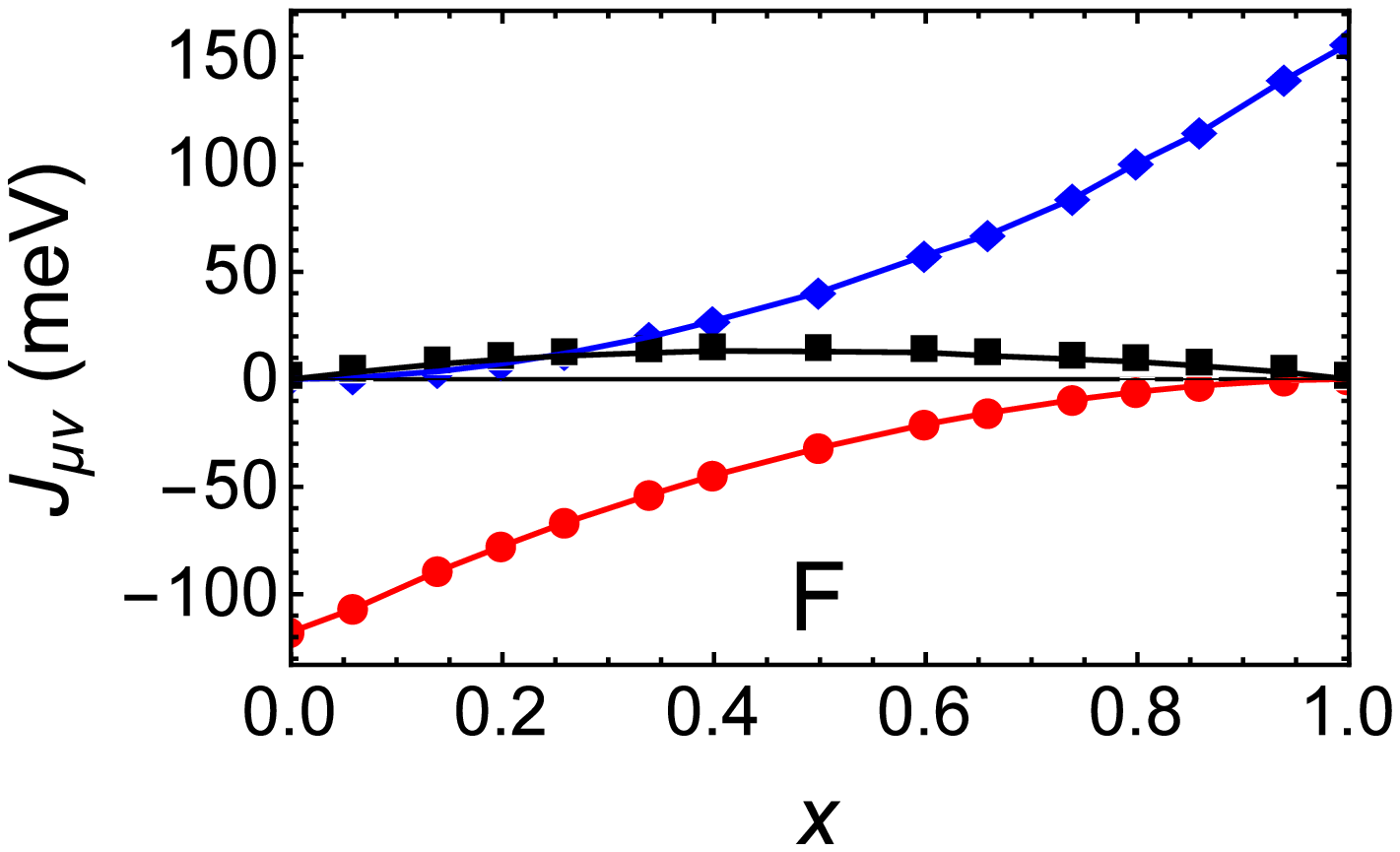}\hfil\includegraphics[height=0.21\textwidth]{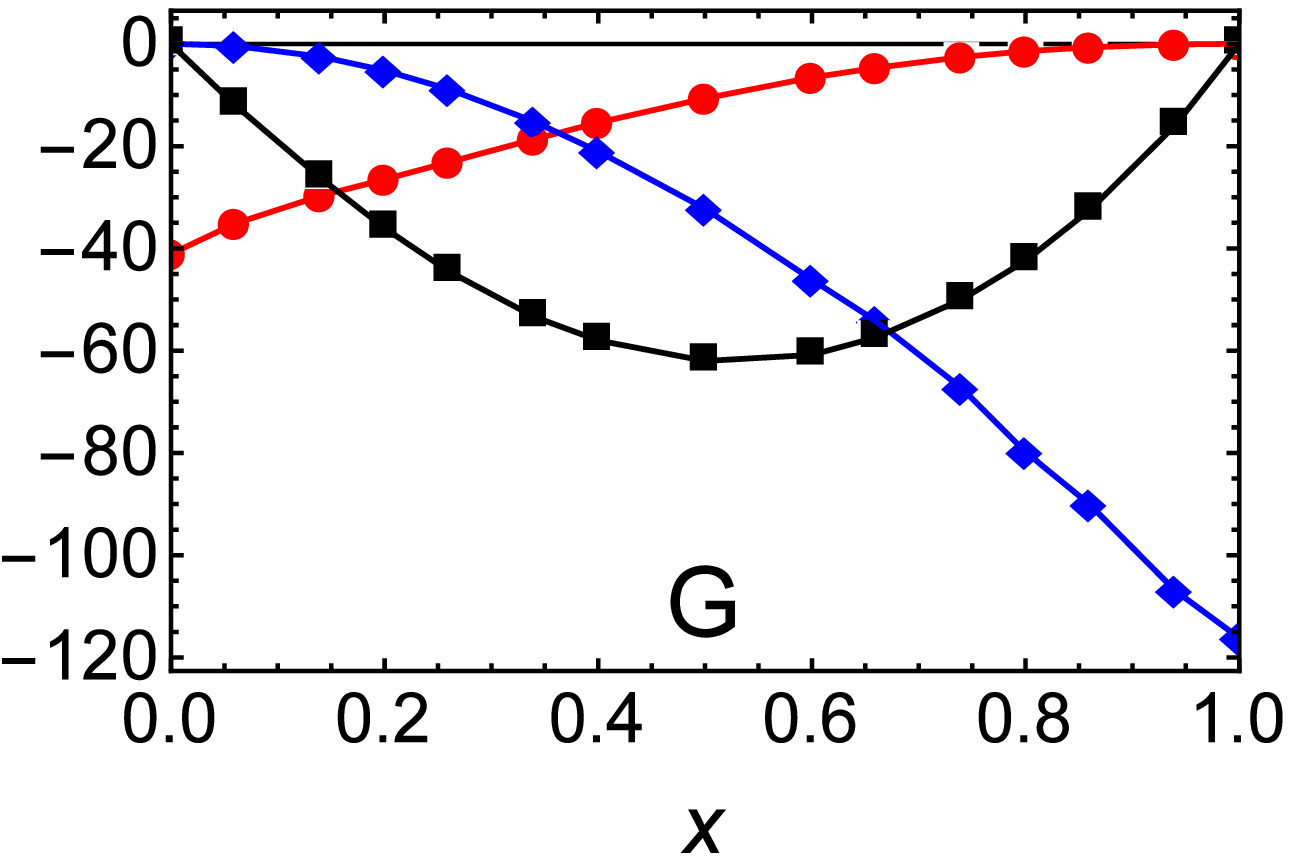}\hfil\includegraphics[height=0.21\textwidth]{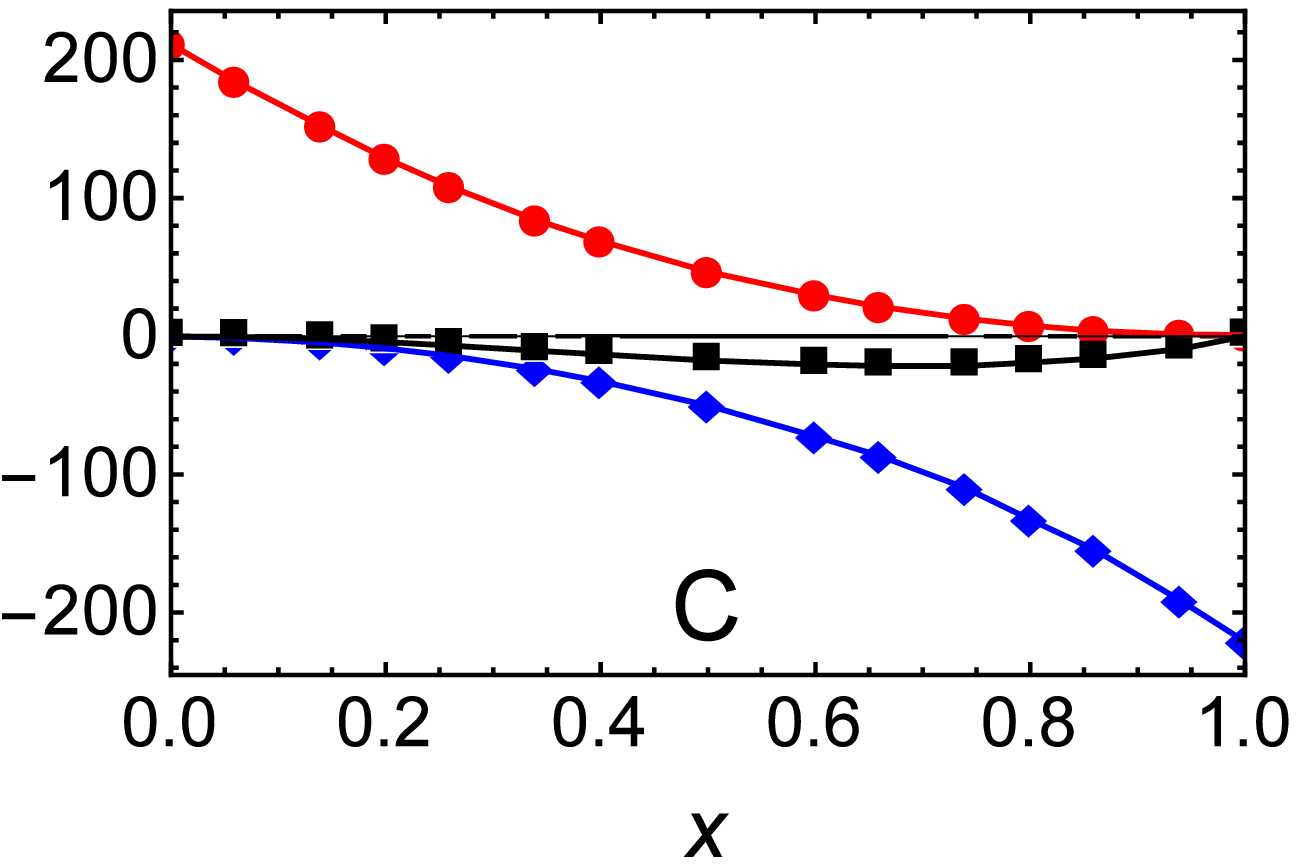}
\caption{Magnetic exchange parameters $J_{\mu\nu}$ in the paramagnetic state corresponding to (F) ferromagnetic ordering, $\mathbf{Q}_F=0$; (G) G-type ordering, $\mathbf{Q}_G=(1,0,1/2)$; (C) C-type ordering, $\mathbf{Q}_C=(1,0,0)$.
Red circles: $J_\mathrm{FeFe}(\mathbf{Q})$; blue diamonds: $J_\mathrm{MnMn}(\mathbf{Q})$; black squares: $J_\mathrm{FeMn}(\mathbf{Q})$.
}
\label{J0}
\end{figure*}

Let us first examine the magnetic interactions in the vicinity of the paramagnetic state. We set up a unit cell for each of the three magnetic orderings and calculate the total energy for about 70 partially ordered VDLM states with $|\alpha_\mathrm{Fe}|$ and $|\alpha_\mathrm{Mn}|$ ranging from 0 to 3. Experimental lattice constants are used at the concentrations reported in Ref.\ \cite{Menshikov}. At each concentration the calculated total magnetic energy $E_{mag}$ (per Fe/Mn atom, referenced from the paramagnetic state) is fitted to even fourth-order polynomials in the reduced magnetizations $m_\mathrm{\mu}=\langle\cos\theta_\mu\rangle$, which are expressed through the fields $\alpha_\mathrm{\nu}$ by the Langevin function. The quadratic part of these polynomials, $E_\mathbf{Q}=\frac12 J_\mathrm{FeFe}(\mathbf{Q})m^2_\mathrm{Fe}+\frac12 J_\mathrm{MnMn}(\mathbf{Q})m^2_\mathrm{Mn}+J_\mathrm{FeMn}(\mathbf{Q})m_\mathrm{Fe}m_\mathrm{Mn}$, defines the component-resolved effective exchange interactions $J_\mathrm{\mu\nu}(\mathbf{Q})$ for the three orderings. Since the total energies are calculated with constraining fields, these exchange parameters are free from the errors associated with the long-wave approximation \cite{Antropov}. The results are shown in Fig.\ \ref{J0}.

\begin{figure*}[htb]
\includegraphics[height=0.21\textwidth]{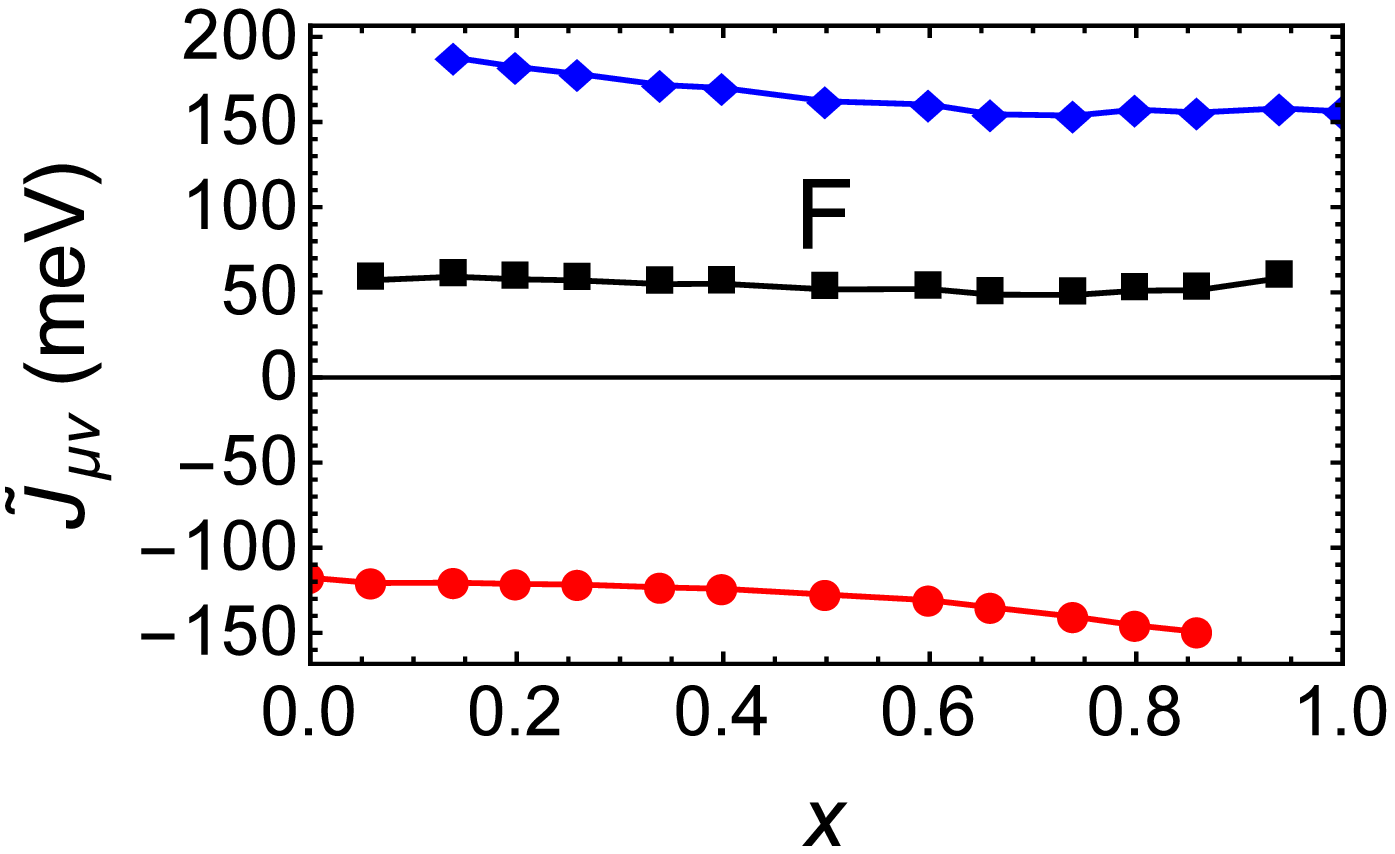}\hfil\includegraphics[height=0.21\textwidth]{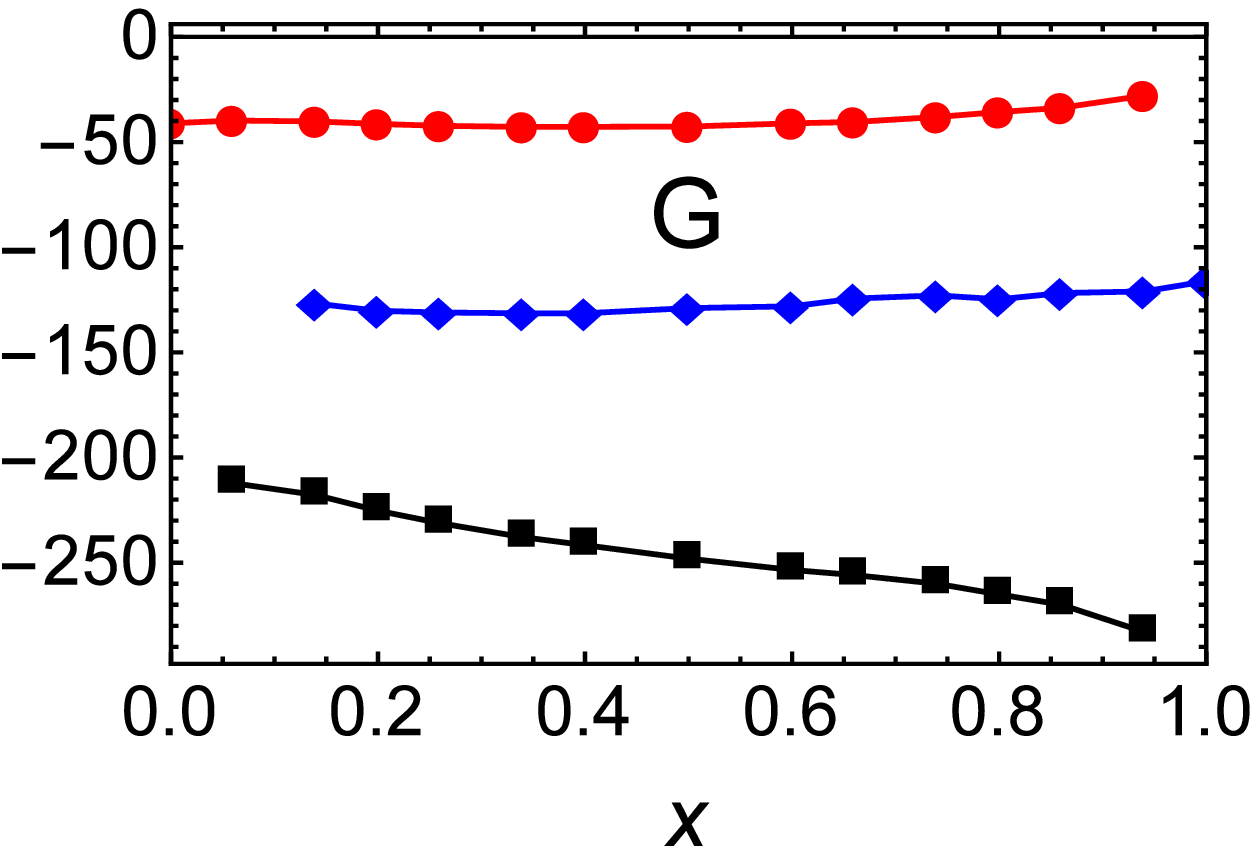}\hfil\includegraphics[height=0.21\textwidth]{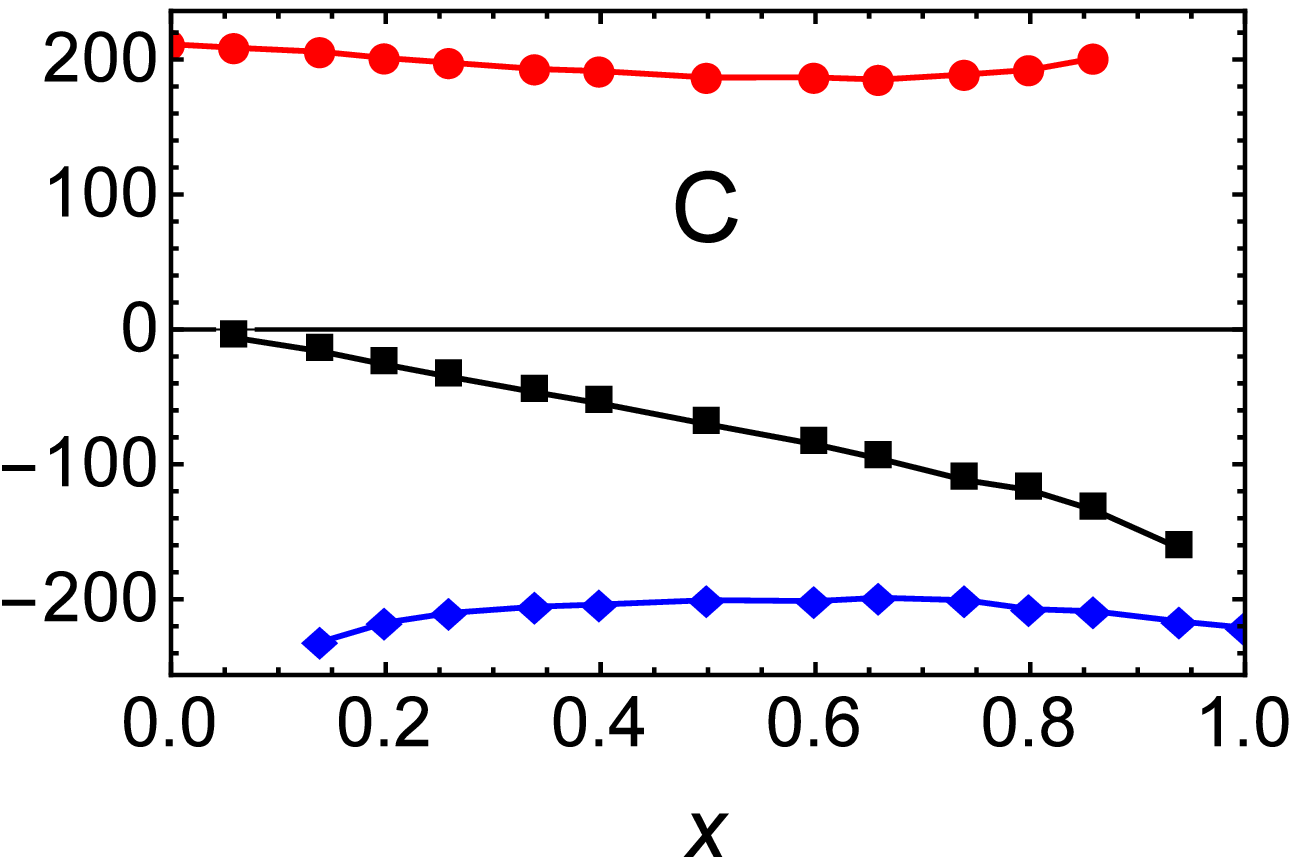}
\caption{Same as in Fig.\ \ref{J0} but normalized by the concentrations: $\tilde J_\mathrm{\mu\nu}(\mathbf{Q})=J_\mathrm{\mu\nu}(\mathbf{Q})/(c_\mu c_\nu)$.}
\label{J0c}
\end{figure*}

The concentration dependence of the parameters $J_\mathrm{\mu\nu}(\mathbf{Q})$ shows that, in agreement with experiment, the F, G, and C-type orderings are favored at the Fe-rich end, in the middle, and at the Mn-rich end, respectively. Further insight can be obtained from the reduced exchange parameters that are normalized by the concentrations, $\tilde J_\mathrm{\mu\nu}(\mathbf{Q})=J_\mathrm{\mu\nu}(\mathbf{Q})/(c_\mu c_\nu)$, shown in Fig.\ \ref{J0c}. These reduced quantities would be concentration-independent in a Heisenberg system with pair exchange parameters $J_{\mu\nu}(\mathbf{R})$ depending only on the distance and the identity of the atoms in a pair. We see that the parameters $\tilde J_\mathrm{\mu\nu}(\mathbf{Q})$ for like components (i.\ e.\ Fe-Fe and Mn-Mn) are almost constant for all ordering vectors, as well as the reduced Fe-Mn coupling for the F ordering. However, the reduced Fe-Mn couplings at the G-type and C-type ordering vectors depend strongly on the concentration, which reflects the effect of band filling on the exchange interaction in metallic systems.

The local spin moments of Fe and Mn increase by about 10\% as $x$ goes from 0 to 1 (for example, from 2.87 to 3.11 $\mu_B$ for Fe and from 3.35 to 3.74 $\mu_B$ for Mn in the paramagnetic phase), and similar variations are observed for different phases and spin directions. The adiabatic approach \cite{DLM} is thus well suited for this system. We also repeated some calculations without the constraining magnetic fields, which is equivalent to making the long-wave approximation \cite{Antropov}, and found that the resulting errors in $J_{\mu\nu}$ for all phases do not exceed 5-7\%.

To further extend the mapping of the magnetic configuration space, we performed noncollinear CPA calculations for the F/G, G/C, and F/C noncollinear $2Q$ phases in the relevant concentration ranges. These calculations are needed to reveal the possible interaction between orderings at different $Q$ in the $2Q$ structures, which can appear in quartic and higher-order interaction terms. A $2Q$ structure is parameterized by two angles, $\theta_\mathrm{Fe}$ and $\theta_\mathrm{Mn}$ that the spin moments of Fe and Mn atoms make with the $z$ axis. (The $z$ and $x$ components of the magnetization order with one or the other of the $Q$ vectors.) Using the symmetries, the accessible space of ($\theta_\mathrm{Fe},\theta_\mathrm{Mn})$ is reduced to the range $0\leq\theta_\mathrm{Fe}\leq\pi/2$, $-\pi\le\theta_\mathrm{Mn}\leq\pi$ with additional $\theta_\mathrm{Mn}\to\pi-\theta_\mathrm{Mn}$ symmetry at $\theta_\mathrm{Fe}=\pi/2$ and $\theta_\mathrm{Mn}\to-\theta_\mathrm{Mn}$ symmetry at $\theta_\mathrm{Fe}=0$. This irreducible domain is covered by a uniform mesh of 38 points.

\begin{figure}[htb]
\includegraphics[height=0.23\textwidth]{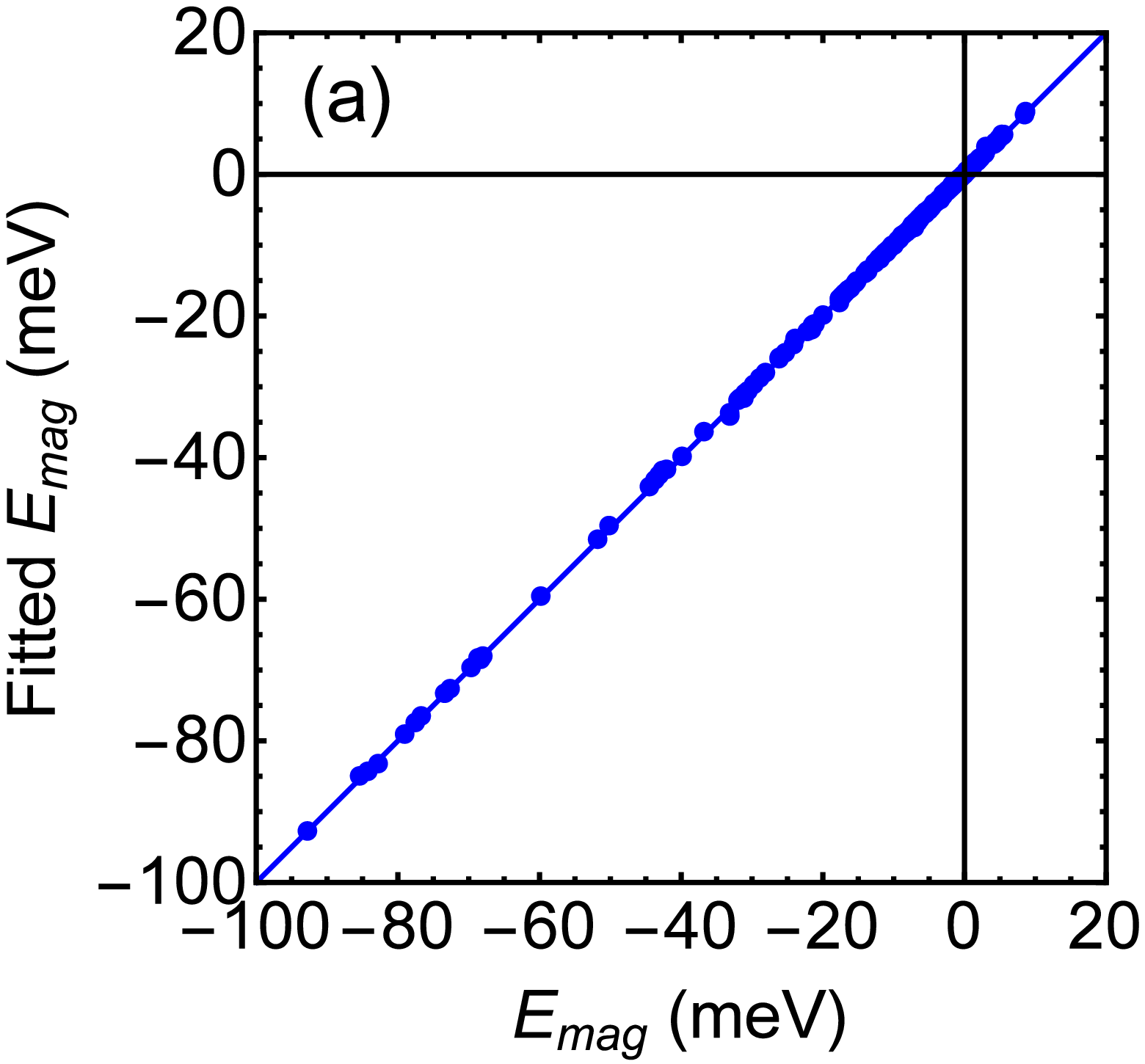}
\hfil
\includegraphics[height=0.23\textwidth]{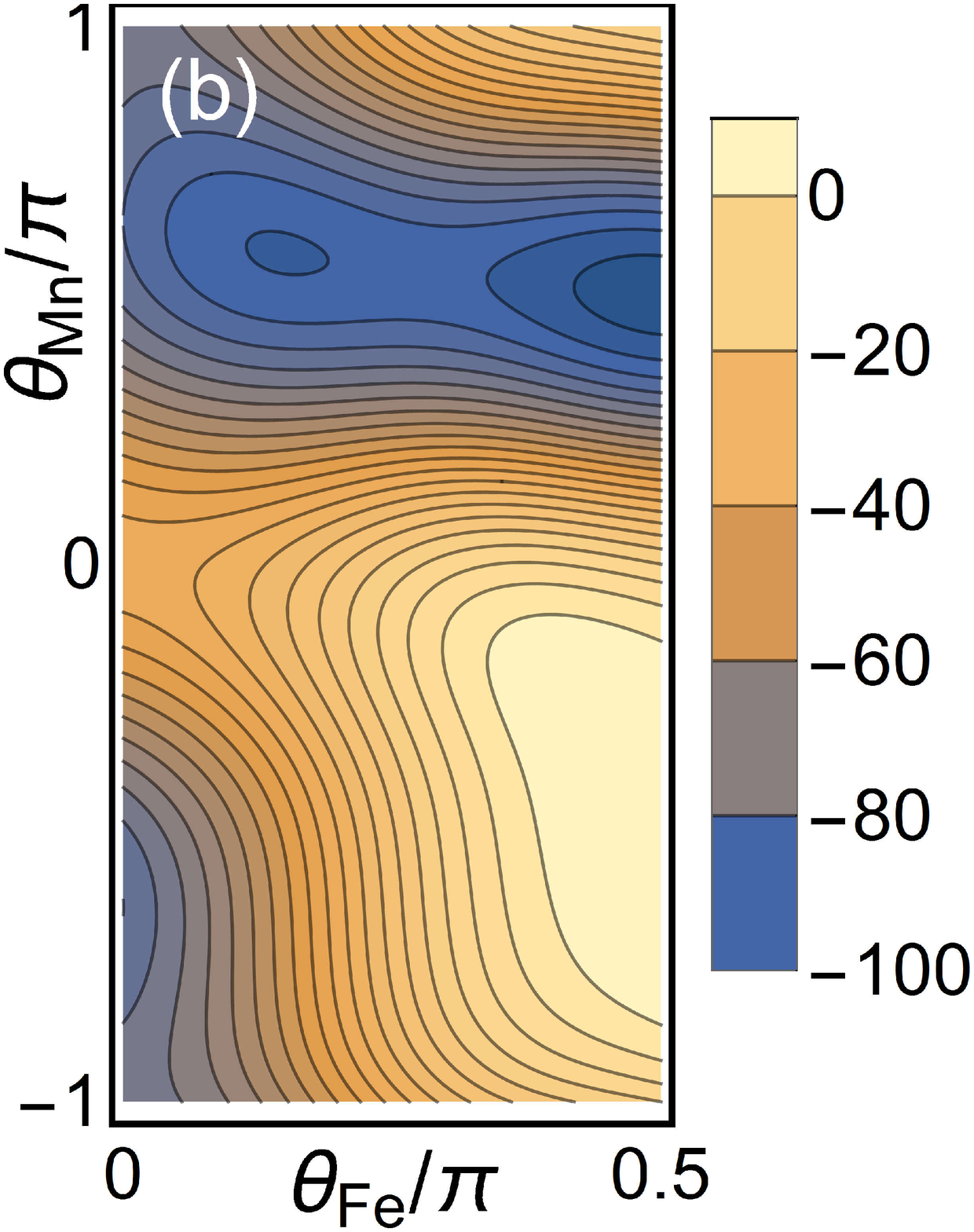}
\caption{(a) Accuracy of the fit at $x=0.26$. (b) The magnetic energy $E_{mag}$ (meV per Fe/Mn atom) in the F/G phase at $x=0.26$. The global minimum at $(\pi/2,\pi/2)$ corresponds to the collinear G-type phase.}
\label{fit}
\end{figure}

For a particular $2Q$ phase (say, F/G) we then combine the data from the separate VDLM calculations for the F and G phases with those from the noncollinear CPA results for the $2Q$ phase and fit the magnetic energy at the given concentration to a polynomial in the order parameters $m_\mathrm{Fe,F}$, $m_\mathrm{Fe,G}$, $m_\mathrm{Mn,F}$, and $m_\mathrm{Mn,G}$. We allowed all symmetry-respecting terms in the polynomials of up to sixth order (see Supplemental Material \cite{sm} for details). The high accuracy of the fits is illustrated in Fig.\ \ref{fit}a for the F/G phase at $x=0.26$; all other fits are of similar accuracy. Fig.\ \ref{fit}b shows the magnetic energy as a function of $\theta_\mathrm{Fe}$, $\theta_\mathrm{Mn}$ in the same phase at $x=0.26$. At this concentration there are two minima: the global one at $(\pi/2,\pi/2)$ corresponding to the collinear G-type phase, and a local one near $(\pi/6,\pi/2)$ corresponding to the $2Q$ phase. At a lower concentration there is a first-order transition where the $2Q$ minimum goes below the G-type minimum.

\begin{figure}[htb]
\includegraphics[width=0.45\textwidth]{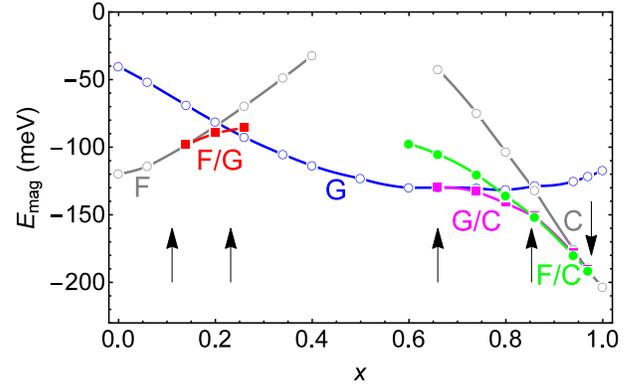}
\caption{Magnetic energies of different phases at zero temperature (per Fe/Mn atom). Arrows show the boundaries between the F, F/G, G, G/C, F/C, and C phases (in the order of increasing $x$).}
\label{ground}
\end{figure}

Using the combined fits for the magnetic energy, we now determine the ground states for all concentrations by choosing the lowest energy of all the competing phases. The results are shown in Fig.\ \ref{ground}. We find that the F phase is stable at $x<0.11$ and the G phase at $0.23<x<0.66$. There is a $2Q$ F/G phase at $0.11<x<0.23$ separated from F by a second-order and from G by a first-order transition. The G/C phase is stable at $0.66<x<0.85$.

Surprisingly, at $x=0.85$ we find a first-order transition from the G/C to the F/C phase. The existence of this first-order transition is in excellent agreement with the observed abrupt drop in the mean magnetic moment at this concentration \cite{Menshikov}. However, since the F component in the low-temperature Mn-rich phase has not been previously identified, the existence of the F/C phase is a prediction that needs to be verified experimentally. The F/C and G/C phases at the Mn-rich end differ essentially in the Fe ordering alone, as the ordering of the Mn spins is almost purely C-type. The energy difference between the F/C and G/C phases reaches about 20 meV per Fe atom near $x=0.95$; it is barely visible in Fig.\ \ref{ground} because of the small Fe concentration.

The first-order transition from F/G to G at $x\approx0.23$ is also in excellent agreement with experiment, while the transition from G to G/C occurs at a larger $x$ compared to experiment, where it is close to $x=0.5$. Note, however, that our calculations are for systems with perfect $3d$/$5d$ ordering, while the order parameter in the experimental samples for $x=0.5$ and $x=0.6$ was 0.79 \cite{Menshikov}.

\begin{figure}[htb]
\includegraphics[width=0.45\textwidth]{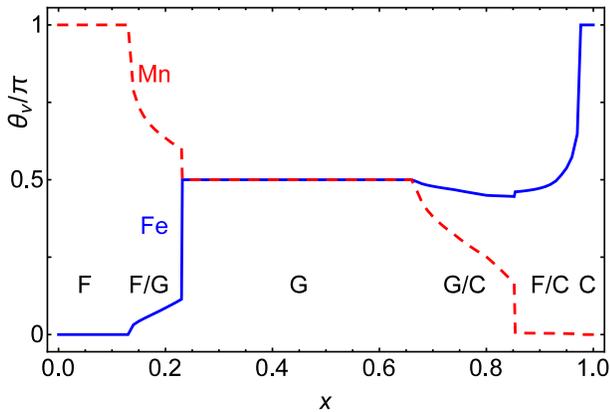}
\caption{Angles $\theta_\nu$ made by the spin moments of Fe and Mn with the $z$ axis at zero temperature. By convention, the G order parameter, as well as the F order parameter in the F/C phase, are assumed to be orthogonal to the $z$ axis; all others are parallel to $z$.}
\label{angles}
\end{figure}

Fig.\ \ref{angles} shows the angles $\theta_\mathrm{Fe}$, $\theta_\mathrm{Mn}$ in the ground states as a function of $x$. (By convention, F and C amplitudes lie along the $z$ and G along the $x$ axis; except for the F/C phase, where F is along $x$.) As could be inferred from Fig.\ \ref{J0}, the spin moments of Fe and Mn are antiparallel in the F phase and parallel in the G phase. First-order phase transitions appear as discontinuous jumps of the angles. Note that in a wide concentration range at the Mn-rich end the Fe spin moments are almost perpendicular to the MnPt host in the ground state, while Mn ordering is almost pure C-type. (A full set of first-principles calculations with interpolated lattice constants was performed with a small 0.01 step in $0.94<x<1$ range to confirm this.) This feature highlights the importance of non-Heisenberg terms in the magnetic energy.

We now turn to the full concentration-temperature phase diagram. We have the fits for the magnetic energy $E_{mag}(\mathbf{m}_\mathrm{Fe},\mathbf{m}_\mathrm{Mn})$, where the $x$ and $z$ components of $\mathbf{m}_\mathrm{Fe}$ and $\mathbf{m}_\mathrm{Mn}$ correspond to the two $Q$ vectors; in the collinear phase one of these components vanishes. $E_{mag}$ is available for discrete concentrations, and we use linear interpolation between them. The entropy is approximated as $S=(1-x)S(m_\mathrm{Fe})+xS(m_\mathrm{Mn})$, where $S(m)$ is the entropy of a classical spin in a Weiss field of such a magnitude that the magnetization is $m$. This corresponds to the mean-field-like distribution function $p_\mathrm{\nu}(\theta')\propto\exp(\alpha_\mathrm{\nu}\cos\theta')$, where $\theta'$ is the angle with respect to the direction of $\mathbf{m}_\mathrm{\nu}$. Given that $m_\mathrm{Fe,F}=m_\mathrm{Fe}\cos\theta_\mathrm{Fe}$ in the F/G phase, etc., we minimize the free energy for each phase with respect to four parameters $\theta_\mathrm{Fe}$, $\theta_\mathrm{Mn}$, $\alpha_\mathrm{Fe}$ and $\alpha_\mathrm{Mn}$.

For a system with purely Heisenberg interaction this scheme is identical to MFA applied to $2Q$ phases \cite{Medvedev}. Therefore, both second-order transitions to the paramagnetic phase (where non-Heisenberg terms have no effect in MFA) and ground-state properties (where only energy is important) are correctly described by this approach. In the intermediate temperature range our scheme can therefore be treated as a thermodynamic interpolation. At such intermediate temperatures we are essentially assuming that the presence of non-Heisenberg terms does not strongly change the distribution functions and that our magnetic energy fits remain valid for partially disordered $2Q$ phases. In most cases, these assumptions are likely consistent with the accuracy of the MFA itself. Significant improvement of the thermodynamic description could be reached by mapping the total energies to a microscopic spin Hamiltonian, but this expensive procedure is beyond the scope of this paper.

\begin{figure}[htb]
\includegraphics[width=0.45\textwidth]{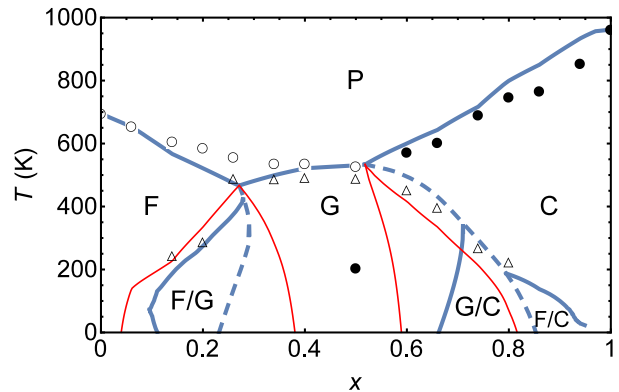}
\caption{Magnetic phase diagram of \femnpt. Temperatures are rescaled by the ratio $T_{exp}(x)/T_{th}(x)$, where $T(x)=(1-x)T_C+x T_N$; $T_C$ and $T_N$ are the ordering temperatures of FePt and MnPt from Ref.\ \onlinecite{Menshikov} ($exp$) or theory ($th$). (MFA gives $T_C=924$ K and $T_N=1670$ K.) Thick (blue) lines: full fit for $E_{mag}$. Thin (red) lines: same fit but with non-Heisenberg terms set to zero. Solid (dashed) lines: second (first) order phase transitions. Symbols: experimental data \cite{Menshikov}.}
\label{PD}
\end{figure}

Fig.\ \ref{PD} shows the MPD obtained both from the full magnetic energy fits and from the same fits truncated at the quadratic (Heisenberg) terms. First-order transitions are shown by dashed lines. The overall structure of the phase diagram agrees well with experiment. The first-order transition from the F/G to the G phase may help explain the peculiarities of the observed temperature-dependent magnetization peaks \cite{Menshikov}. Indeed, there is a concentration range from $x=0.23$ to $x\approx0.28$ where the ground-state G-type ordering turns into F/G and then to F on heating. The physics is complicated by configurational disorder, which may lead to the formation of small Fe-rich clusters at elevated temperatures \cite{Menshikov}.

If only Heisenberg terms are kept in the magnetic energy, the transitions into the paramagnetic phase remain unchanged. However, all first-order transitions turn into second-order; the F/G to G transition is shifted to much larger concentrations, and the F/C phase disappears completely. Thus, in order to describe the observed first-order transitions at $x=0.23$ and $x=0.85$ it is important to take the non-Heisenberg interaction terms into account.
The F/C phase disorders at rather low temperatures in our description, although its stability may be underestimated in our thermodynamic scheme. The non-zero magnetization in this phase should facilitate an easy experimental verification of its existence.

To conclude, we developed a computational tool based on a combination of first-principles calculations that is capable of describing a complicated $c$-$T$ magnetic phase diagram of a metallic alloy with competing interactions. Its application to the \femnpt\ system produced a detailed interpretation of the experimental phase diagram and also predicted the existence of a previously unknown low-temperature magnetic phase on the Mn-rich end. The correct first-order transitions and the Mn-rich phase are only captured if non-Heisenberg terms are included in the magnetic energy, showing the limitations of the conventional approach based on the Heisenberg-model Hamiltonian. The wide applicability and predictive power of this approach makes it useful for the design of magnetic materials with desired properties.

\begin{acknowledgments}

The work at UNL was supported by the National Science Foundation through Grant No.\ DMR-1308751 and the Nebraska MRSEC (DMR-0820521) and was performed utilizing the Holland Computing Center of the University of Nebraska. Work at Ames Lab is supported in part by the Critical Materials Institute, an Energy Innovation Hub funded by the US DOE and by the Office of Basic Energy Science, Division of Materials Science and Engineering. Ames Laboratory is operated for the US DOE by Iowa State University under Contract No.\ DE-AC02-07CH11358.

\end{acknowledgments}

\balancecolsandclearpage

\section{Supplemental Material}

\setcounter{figure}{0}
\makeatletter
\renewcommand{\thefigure}{S\arabic{figure}}
\renewcommand{\bibnumfmt}[1]{[S#1]}
\renewcommand{\citenumfont}[1]{S#1}
\makeatother

In its basic features, our implementation \cite{Fe8Ns} of the coherent potential approximation (CPA) follows the formulation \cite{Tureks} based on iterating the ``coherent interactor'' matrix $\Omega$ for sites with chemical and/or spin disorder, which is defined as
\begin{equation}
\Omega(z) = \mathcal{P}(z)-g^{-1}(z),
\end{equation}
where $z$ is the complex energy, $\mathcal{P}$ is the coherent potential and $g$ the on-site auxiliary Green's function. The latter is computed as the on-site block of the $\mathbf{k}$-integrated crystal Green's function $g(\mathbf{k})=[\mathcal{P}_B\delta_{BB'}-S_{BB'}(\mathbf{k})]^{-1}$, where $B$ and $B'$ denote basis sites within the unit cell, and $S$ is the structure constant matrix of the tight-binding linear muffin-tin orbital (LMTO) method. The coherent potential is in turn calculated from the LMTO potential parameters $P_a$ and the $\Omega$ matrix:
\begin{equation}
\mathcal{P} = \biggl[\sum_a p_a (P_a-\Omega)^{-1} \biggr]^{-1} + \Omega
\label{P}
\end{equation}
where $p_a$ is the probability to find an atom of the component $a$ on the given site.

In the vector disordered local moment method (VDLM), different directions of the local magnetic moments are treated as different alloy components. The component index $a$ in Eq.\ (\ref{P}) then combines the index of the chemical component $\mu$ and the orientation of the local moment $\hat n_\mu$, and the probabilities become $p_a = c_\mu p_\mu(\hat n_\mu)$, where $c_\mu$ is the concentration of the component $\mu$ on the given site, and $p_\mu(\hat n_\mu)$ is the distribution function for the spin orientations.

The LMTO potential parameter matrix for each spin direction is initially computed in the reference frame in which the polar axis is aligned with the direction $\hat b$ of the total effective magnetic field for this direction (more on this below). Let us denoting this matrix $\bar P_a$. To obtain $P_a$ in Eq.\ (\ref{P}), we need to rotate $\bar P_a$ to the global reference frame:
\begin{equation}
P_a = U(\hat b) \bar P_a U^{-1}(\hat b)
\end{equation}
where $U(\hat b)$ is the rotation operator that rotates the $\hat z$ vector to $\hat b$. In this work spin-orbit coupling is neglected. Therefore, $\bar P_a$ is spin-diagonal and the rotation $U(\hat b)$ only affects the spin indices. If spin-orbit coupling is included \cite{Fe2Bs}, $U(\hat b)$ acts on both spin and orbital indices and is generated by the total angular momentum operator $\hat J$.

The summation over $a$ in Eq.\ (\ref{P}) involves a sum over chemical components and an integration over the sphere for each component treated within the VDLM scheme. For our present purposes the VDLM scheme is implemented for partially ordered magnetic states with a global axial symmetry and without spin-orbit coupling. This symmetry implies that the LMTO potential matrices $\bar P_a$ and the spin distribution functions $p_\mu(\hat n_\mu)$ depend only on the chemical identity $\mu$ and the polar angle $\theta_\mu$ with respect to the magnetic symmetry axis. In this special case the integral over the azimuthal angle in Eq.\ \ref{P} has a very simple effect: it eliminates all spin off-diagonal elements and leaves spin-diagonal elements (which do not depend on $\phi$) unchanged. The remaining integration over the polar angle $\theta_\mu$ is discretized using the 16-point Gauss-Legendre quadrature. Thus, formally we are dealing with 32 CPA components on the Fe/Mn sites corresponding to 16 values of the polar angle for Fe and Mn.

The spin distribution functions for partially ordered VDLM states are assumed to have the Curie-Weiss form $p_{\mu}(\theta)\propto\exp(\alpha_{\mu}\cos\theta)$. The factors $\alpha_{\mu}$ are treated as variational parameters in the free energy minimization.

\begin{figure*}[hbt]
\centering
\includegraphics[width=0.85\textwidth]{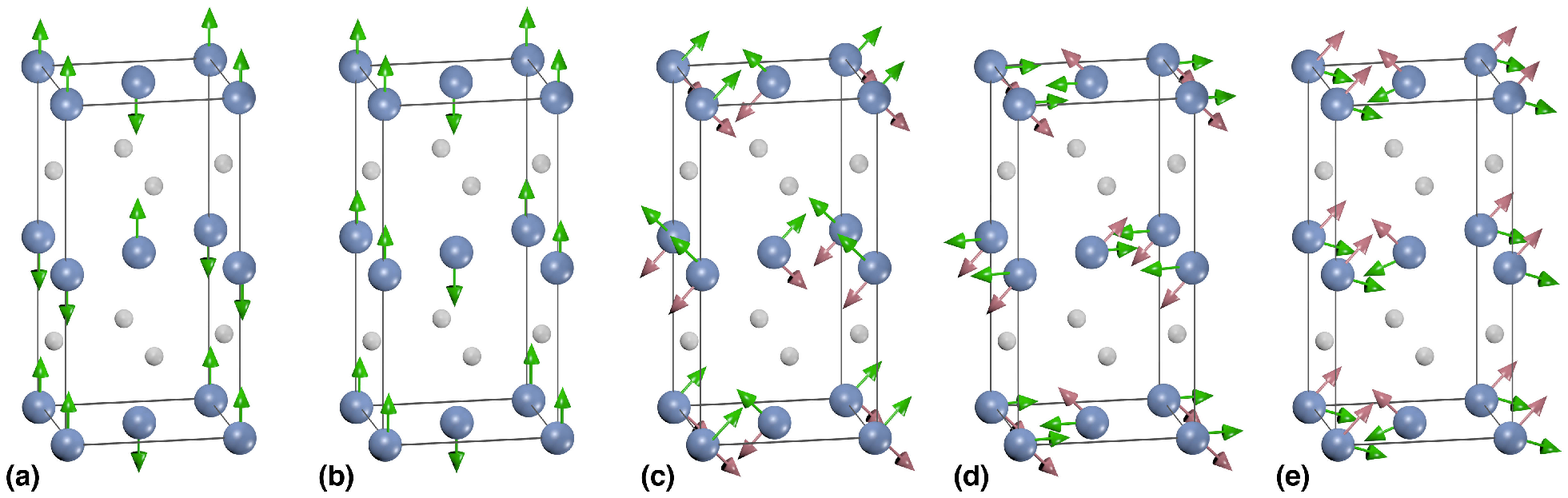}
\vskip3ex
\caption{Magnetic structures: (a) G-type, (b) C-type, (c) F/G, (d) G/C, (e) F/C. Large blue (small gray) spheres show Fe/Mn (Pt) sites. Green and red arrows in panels (c)-(d) show the spins of Fe and Mn atoms, which have similar ordering patterns but are not parallel to each other. Since spin-orbit coupling is disregarded, only relative angles between the spins are important.}
\label{structures}
\end{figure*}

In addition to VDLM states, in the construction of the energy functional we also use CPA calculations for the $2Q$ phases (F/G, G/C, and F/C) with definite (non-random) spin directions. These calculations are essential to capture the higher-order interactions between the order parameters corresponding to different $Q$ vectors; indeed, axially-symmetric VDLM calculations represent partially ordered states with only one ordering $Q$ vector. In these calculations, the sum over $a$ in Eq.\ (\ref{P}) has only two terms for Fe and Mn with their spins pointing in certain directions within the $xOz$ plane (as is sufficient for the $2Q$ states). Of course, the spin off-diagonal terms are retained in Eq.\ (\ref{P}), whereby $\mathcal P$, $\Omega$, and $g$ also become spin off-diagonal. The magnetic configuration in a $2Q$ phase can be written as $\mathbf{m}_{i}^{\mu} = \mathbf{m}_{1}^{\mu} \exp(i\mathbf{Q}_1 \mathbf{R}_i)+\mathbf{m}_{2}^{\mu} \exp(i\mathbf{Q}_2 \mathbf{R}_i)$ with orthogonal order parameters ($\mathbf{m}_{1}^{\mu}\mathbf{m}_{2}^{\nu}=0$), where $\mathbf{R}_i$ is the coordinate of site $i$. Because of the orthogonality, there are four real order parameters in a $2Q$ phase, which can be simply denoted as ${m}^{\mu}_{1}$ and ${m}^{\mu}_{2}$. $\mathbf{m}_{i}^{\mu}$ represents the averaged local moment of an atom of component $\mu$ appearing at lattice site $i$. The site magnetizations are defined as $m_\mathrm{\mu}=\langle\cos\theta_\mu\rangle$, i.\ e.\ the dependence of the local moment on its orientation is neglected \emph{in this definition}. This choice does impair the accuracy of the calculations, because the fitting of the calculated total energies and the calculation of the magnetic entropy are consistent with this specific definition of the order parameters.
The relevant ordering wave vectors are $\mathbf{Q}_F=(0,0,0)$, $\mathbf{Q}_C=(1,0,0)$, and $\mathbf{Q}_G=(1,0,1/2)$ in units of $2\pi/a$ (or $2\pi/c$ for the $z$ component). All these wave vectors are commensurate with a tetragonal $a\times a\times2c$ unit cell, which is therefore used for the calculation of the total energies in all magnetic phases. (The use of the same unit cell improves the accuracy of the calculations.)
The magnetic structures of the considered $2Q$ states are shown in Fig.~\ref{structures} along with the $G$-type and $C$-type orderings (the ferromagnetic state is not shown).

Using the symmetries $\theta_\mu\to\pi-\theta_\mu$ and $\theta_\mu\to-\theta_\mu$ (applied to all $\mu$ at the same time), the accessible space of ($\theta_\mathrm{Fe},\theta_\mathrm{Mn})$ is reduced to the range $0\leq\theta_\mathrm{Fe}\leq\pi/2$, $-\pi\le\theta_\mathrm{Mn}\leq\pi$ with additional $\theta_\mathrm{Mn}\to\pi-\theta_\mathrm{Mn}$ symmetry at $\theta_\mathrm{Fe}=\pi/2$ and $\theta_\mathrm{Mn}\to-\theta_\mathrm{Mn}$ symmetry at $\theta_\mathrm{Fe}=0$. This irreducible domain is covered by a uniform square mesh of 38 inequivalent points.

The LMTO charges are computed in the usual way from the conditionally-averaged physical Green's function \cite{Tureks} $G_a = \lambda_a + \mu_a g_a \mu_a$ in the reference frame in which the polar axis is aligned with the prescribed spin direction $\hat n$. Here $\lambda_a$ and $\mu_a$ are the spin-diagonal potential parameters related to $\bar P_a$, and $g_a=[\bar P_a-U^{-1}(\hat n) \Omega U(\hat n)]^{-1}$. We emphasize that the LMTO charges and potentials are different for different orientations of the local moment, and self-consistency is achieved independently for all of them. Since the coherent interactor $\Omega$, which includes the magnetic coupling to the rest of the crystal, and $P_a$, describing the local exchange-correlation field, are diagonal in different reference frames, the magnetic moment obtained from $G_a$ deviates from the direction $\hat b$ of the total effective magnetic field. If the transverse component of this output magnetic moment is discarded, the obtained magnetic state is not self-consistent in the sense of the density functional theory (DFT). In order to remedy this problem, constraining transverse magnetic fields need to be introduced \cite{Dederichss,DLMs}. We implement these fields in the form suggested in Ref.\ \onlinecite{Stockss}. Specifically, the constraining field for each atom is assumed to have the same radial dependence as the exchange-correlation field on the same atom. The total effective magnetic field is then collinear within the atomic sphere, oriented at an angle $\delta$ to the exchange-correlation field, and rescaled by a factor $\sqrt{1+\delta^2}$ \cite{Stockss}. At each iteration toward self-consistency, the misalignment angle $\delta$ is adjusted to eliminate the deviation of the magnetic moment from the prescribed direction $\hat n$. Thus, at self-consistency, the local moment is collinear with the exchange-correlation field direction $\hat n$, while the total effective magnetic field is rotated by an angle $\delta$ with respect to $\hat n$.

For the $2Q$ states there is an additional complication that the orientations of the induced local moments on the Pt atoms are not known in advance. These orientations are also updated at each iteration using the output directions of the induced magnetic moments, guaranteeing full DFT self-consistency in the final state.

As explained in the main text, the VDLM and noncollinear CPA calculations were performed for a number of concentrations with experimental lattice parameters. The calculated (about 200) total energies for each candidate $2Q$ phase at the given concentration are then fitted to a polynomial in the four order parameters $m^{\mu}_{1}$, $m^{\mu}_{2}$ (where 1 and 2 label the ordering types, e.g. F and G for phase F/G): $P(m^{\mu}_{1},m^{\mu}_{2})=P_{1}(m^{\mu}_{1})+P_{2}(m^{\mu}_{2})+P_{12}(m^{\mu}_{1},m^{\mu}_{2})$. The polynomials $P_1$ and $P_2$ include all even terms up to sixth order, and $P_{12}$ includes all products of second-order monomials in $m^{\mu}_{1}$ and $m^{\mu}_{2}$ (lower-order terms in $P_{12}$ are forbidden by symmetry). This fit also covered the two $1Q$ phases, which are obtained at $m^{\mu}_{1}=0$ or $m^{\mu}_{2}=0$.

Fig.\ 3 of the main text illustrates the quality of the fit and its energy profile for fully ordered F/G phase at one selected concentration $x=0.26$. The actual expression \cite{request} is the following ($x_j\equiv m^\mathrm{Fe}_j$, $y_j\equiv m^\mathrm{Mn}_j$):
\begin{widetext}
\begin{align*}
P_F(x_F,y_F)=&-64.78 x_F^2+11.94 x_F y_F+11.96 y_F^2\\
&-11.79 x_F^4+0.85 x_F^3 y_F+5.66 x_F^2 y_F^2+1.79 x_F y_F^3+1.12 y_F^4\\
&+14.39 x_F^6+7.75 x_F^5 y_F-0.69 x_F^4 y_F^2-2.25 x_F^3 y_F^3-5.48 x_F^2 y_F^4-2.08 x_F y_F^5-2.00 y_F^6\\
P_G(x_G,y_G)=&-25.25 x_G^2-45.15 x_G y_G-9.18 y_G^2\\
&+13.02 x_G^4+2.16 x_G^3 y_G-5.77 x_G^2 y_G^2-3.16 x_G y_G^3+1.14 y_G^4\\
&-17.68 x_G^6-5.56 x_G^5 y_G+0.82 x_G^4 y_G^2-2.76 x_G^3 y_G^3+2.22 x_G^2 y_G^4+3.65 x_G y_G^5-1.2 y_G^6\\
P_{FG}(x_F,y_F,x_G,y_G)=&+23.09 x_F^2 x_G^2+0.92 x_F^2 x_G y_G-1.82 x_F^2 y_G^2+5.72 x_F y_F x_G^2 +5.42 x_F y_F x_G y_G-4.5 x_F y_F y_G^2\\
&+9.81 x_G^2 y_F^2+1.61 y_F^2 x_G y_G-8.61 y_F^2 y_G^2\\
\end{align*}
\end{widetext}

The polynomial $P_{12}$ describes interaction between the order parameters at two different $Q$ vectors. Information needed to fit this polynomial is only contained in the noncollinear CPA results for the given $2Q$ phase. ($P_{12}$ vanishes in the $1Q$ phases described by the VDLM method.) Since these calculations (contrary to VDLM ones) are done only for magnetically-ordered phases, certain higher-order terms in $P_{12}$ would be linearly dependent within the available dataset and could not be distinguished. Indeed, in a fully-ordered $2Q$ phase we have $(m^\mu_1)^2+(m^\mu_2)^2=1$, and terms like $x_1^2x_2^2$, $x_1^2x_2^4$, and $x_1^4x_2^2$ are linearly dependent. Therefore, for most concentrations we restricted $P_{12}$ to fourth-order terms, which are expected to be the most important. However, at the Mn-rich end the interaction of the Fe spins with the Mn host has a particularly strong non-Heisenberg character in the F/C phase, making the fourth-order fit for $P_{12}$ insufficient. Therefore, at $x\geq0.8$ we added sixth-order terms to $P_\mathrm{FC}$ that are second-order in F and fourth-order in C order parameters (such as $x_F^2x_C^4$, etc.). In addition, to provide a consistent fit for both G/C and F/C phases, in this concentration range we combined all input data to obtain a single fit for the $3Q$ structure involving all three order parameters. (Possible interaction between the F and G order parameters, which is irrelevant in this region, was left undetermined.)


\begin{thebibliography}{99}
\bibitem{request} The expressions for other concentrations and phases are available from the authors on request.
\end{thebibliography}
\end{document}